\documentclass[english,aps,prl, twocolumn, footinbib, showpacs, showkeys, superscriptaddress,longbibliography]{revtex4-1}
\usepackage[T1]{fontenc}
\usepackage{color}
\usepackage{babel}
\usepackage{bm}
\usepackage{amsmath}
\usepackage{amssymb}
\usepackage{graphicx}
\usepackage[unicode=true,
 bookmarks=true,bookmarksnumbered=false,bookmarksopen=false,
 breaklinks=true,pdfborder={0 0 1},backref=false,colorlinks=true]
 {hyperref}
\hypersetup{
 pdfauthor={Konschelle - Bergeret - Tokatly},
 citecolor=blue,linkcolor=magenta,urlcolor=blue}
\usepackage{breakurl}

\makeatletter
\usepackage{eucal}

\DeclareMathOperator{\Tr}{Tr}

\DeclareMathOperator{\Pexp}{Pexp}

\makeatother

\begin{document}

\title{Semiclassical quantization of spinning quasiparticles in ballistic
Josephson junctions}

\author{Fran\c{c}ois Konschelle}

\affiliation{Centro de F\'{\i}sica de Materiales (CFM-MPC), Centro Mixto CSIC-UPV/EHU,
Manuel de Lardizabal 5, E-20018 San Sebasti\'{a}n, Spain}

\author{F. Sebasti\'{a}n Bergeret}

\affiliation{Centro de F\'{\i}sica de Materiales (CFM-MPC), Centro Mixto CSIC-UPV/EHU,
Manuel de Lardizabal 4, E-20018 San Sebasti\'{a}n, Spain}

\affiliation{Donostia International Physics Center (DIPC), Manuel de Lardizabal
5, E-20018 San Sebasti\'{a}n, Spain}

\author{Ilya V. Tokatly}

\affiliation{Nano-Bio Spectroscopy group, Departamento F\'{\i}sica de Materiales,
Universidad del Pa\'{\i}s Vasco, Av. Tolosa 72, E-20018 San Sebasti\'{a}n,
Spain }

\affiliation{IKERBASQUE, Basque Foundation for Science, E-48011 Bilbao, Spain}
\begin{abstract}
A Josephson junction made of a generic magnetic material sandwiched
between two conventional superconductors is studied in the ballistic
semi-classic limit. The spectrum of Andreev bound states is obtained
from the single-valuedness of a particle-hole spinor over closed orbits
generated by electron-hole reflections at the interfaces between superconducting
and normal materials. The semiclassical quantization condition is
shown to depend only on the angle mismatch between initial and final
spin directions along such closed trajectories. For the demonstration,
an Andreev-Wilson loop in the composite position/particle-hole/spin
space is constructed, and shown to depend on only two parameters,
namely a magnetic phase shift and a local precession axis for the
spin. The details of the Andreev-Wilson loop can be extracted via
measuring the spin-resolved density of states. A Josephson junction
can thus be viewed as an analog computer of closed-path-ordered exponentials.
\end{abstract}



\date{\today}

\maketitle
In the last years the spin-orbit and spin-splitting effects in superconducting
heterostructures \cite{buzdin.2005_RMP,Bergeret2005} are receiving
a great deal of attention in the context of an emerging superconducting
spintronics \cite{Eschrig2011,Linder2015} and in connection with
possible realizations of Majorana bound states in nanowires \cite{franz2013majorana}.
A Josephson junction with a magneto-active normal bridge exemplify
a prototype structure hosting such kind of spin interactions. The
physics of superconductor/normal metal/superconductor (S/N/S) ballistic
Josephson junctions is mainly determined by the so called Andreev
bound states (ABS) localized in the N-region. These states, which
carry a significant fraction of the Josephson supercurrent \cite{beenakker.RMP.1997,Furusaki1991},
have been extensively studied in ballistic superconducting point contacts
\cite{DellaRocca2007,Bretheau2013a,Janvier2015}. 

\begin{figure}[b]
\includegraphics[width=0.95\columnwidth]{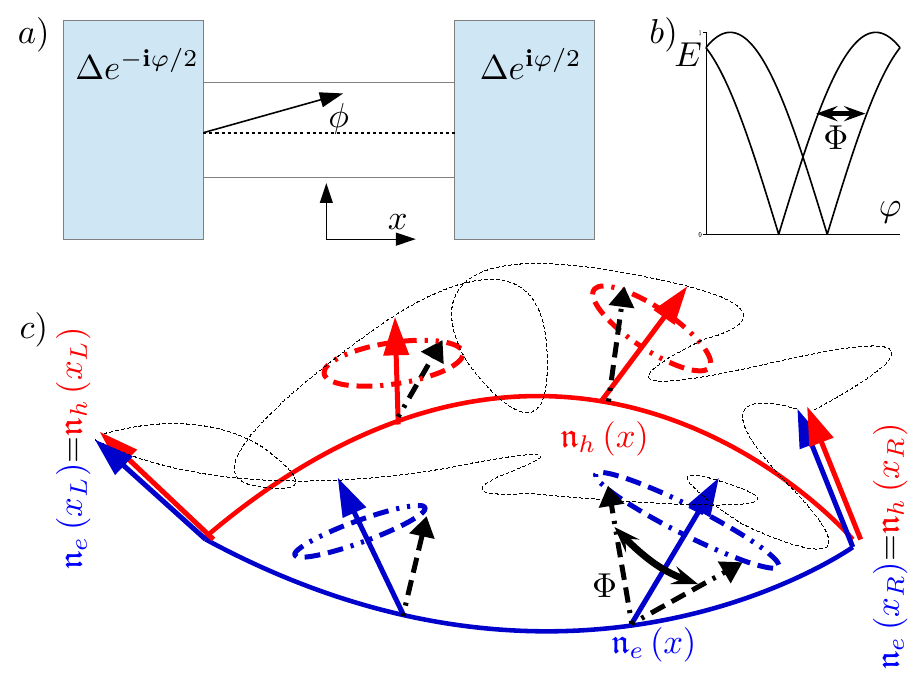}

\caption{\label{fig:AW-loop}a) Sketch of the Josephson junction, with $\phi$
the Fermi angle and $\varphi$ the phase-difference. b) Sketch of
the energy of the Andreev bound states, versus $\varphi$, spin-split
by $\Phi$. c) The spin structure of the Andreev-Wilson loop that
starts at point $x$. The local $\boldsymbol{\mathfrak{n}}_{e}$ (blue)
vector evolves from the left to the right interfaces, where it is
equal to $\boldsymbol{\mathfrak{n}}_{h}$ (red) which propagates in
the opposite direction ($\boldsymbol{\mathfrak{n}}_{e}\left(x_{L,R}\right)=\boldsymbol{\mathfrak{n}}_{h}\left(x_{L,R}\right)$).
The spin direction (black dotted arrows) precess around the local
vectors $\boldsymbol{\mathfrak{n}}_{e,h}$ at a constant latitude
(red and blue dotted projective circles). After completing the loop,
the spin rotates an angle $\Phi$ between the initial and final states.
This angle\textcolor{red}{{} }\textcolor{black}{determines the phase
shift} between the two spin-split Andreev bound states (see b).}
\end{figure}

Theoretically the quantization of states trapped in some classically
allowed region can be understood from the Bohr-Sommerfeld quantization
rule \cite{b.abrikosov,Messiah1995} which requires the phase accumulated
along a closed classical trajectory to be a multiple of $2\pi$. In
a ballistic S/N/S junction the trapping in the N-region occurs due
to Andreev reflections with conversion of the incident electron to
the reflected hole and vice versa at the S/N interfaces \cite{Andreev1964}.
Each Andreev reflection brings a phase shift $\theta\left(E\right)=\arccos\left(E/\Delta\right)$,
where $E<\Delta$ is the energy measured with respect to the Fermi
level \footnote{In the strict semiclassical limit, i.e. $E\rightarrow0$, $\theta\left(0\right)=\pi/2$
is the usual Maslov index at a turning point \cite{Duncan2002}}. The classical loop trajectory is now defined in the space composed
of the position and particle-hole subspaces. In the position subspace
the electron and the reflected hole accumulate the phase equal to
$2EL/v$, where $L$ is the distance between the S electrodes and
$v$ is the component of the velocity perpendicular to the junction
plane. From the two Andreev reflections (shifts in the particle-hole
subspace) the phase acquires the contribution $2\theta\left(E\right)\pm\varphi$,
depending on the propagating direction, where $\varphi$ is the phase
difference between the two S-electrodes (see Fig.\ref{fig:AW-loop}),
\cite{Beenakker1991}. Hence the quantization condition for ABS reads:
$2E_{n}L/|v|-2\theta\left(E_{n}\right)+{\rm sgn}(v)\varphi=2n\pi$.
The spin-orbit coupling (SOC) and spin-splitting (exchange or Zeeman),
possibly textured fields in a magnetic material, generate precession
of the electron and hole spins, which should modify the properties
of ABS. How the semiclassical condition is modified in the presence
of generic spin-dependent fields is an open question we address in
this letter. 

We identify an additional phase shift $\Phi$ originating from the
spin precession generated by an effective magnetic field in the N-region,
see Eq. \eqref{eq:BS-spectrum}. This precession preserves, as in
the non-superconducting case \cite{Keppeler2002a,Keppeler2002}, the
latitude with respect to the local spin quantization axis ${\bf \mathfrak{n}}$,
which obeys a classical equation {[}Eq. \eqref{eq:ne-nh}{]}. We first
derive the modified quantization condition, determine the subgap spectrum
of a ballistic S/N/S junction and finally demonstrate by solving the
quasiclassical Eilenberger equation how $\Phi$ and ${\bf \mathfrak{n}}$
enter the expressions of other physical quantities like the Josephson
current or the spin resolved local density of states. Our results
generalize the quasiclasical theory of spinning electrons described
by Dirac and/or Pauli equations \cite{Keppeler2002a,Keppeler2002}
to the case of quasiclassical motion of Bogoliubov quasiparticles
in superconducting structures. In the normal state a non-adiabatic
spin precession of electrons moving along cyclotron orbits is revealed
experimentally in anomalous Shubnikov -- de Haas oscillations of magnetoresistance
\cite{KepWin2002}. Here we demonstrate that all details of highly
nontrivial spin dynamics of bogolons forming ABS can be extracted
from the observable properties of Josephson junctions.

We consider the semiclassical Bogoliubov-deGennes (BdG) bispinor wave
function $(u,v)=e^{\mathbf{i}\mathbf{k}\cdot{\bf r}}(\psi,\text{\ensuremath{\chi}})$,
where $\left|{\bf k}\right|=k_{F}$ is the Fermi momentum, and the
electron $\psi$ and hole $\chi$ spinors are slowly varying on the
scale of $k_{F}^{-1}$ \footnote{We assume the validity of the semiclassical approach. This means that
we assume that all lengths and involved in the problem are larger
than the Fermi wave length and all energies, in particular the SOC
and Zeeman splitting, are smaller than the Fermi energy.}. The behavior of the wave function in the presence of an effective,
coordinate- and velocity-dependent, magnetic field $\boldsymbol{B}(\bm{v},x)$,
which couples to the electron and hole spins and describes generic
SOC and exchange/Zeeman spin splitting, is governed by the following
BdG equations,
\begin{align}
-\mathbf{i}\boldsymbol{v\cdot\nabla}\psi+\boldsymbol{B}\left(\boldsymbol{v},x\right)\boldsymbol{\cdot\sigma}\psi+\Delta\chi & =E\psi\nonumber \\
\mathbf{i}\boldsymbol{v\cdot\nabla}\chi+\boldsymbol{B}\left(-\boldsymbol{v},x\right)\boldsymbol{\cdot\sigma}\chi+\Delta^{\ast}\psi & =E\chi\;.\label{eq:BdG-1-1}
\end{align}
 We do not impose any restriction on the $\bm{v}$- or $x$- dependence
of $\boldsymbol{B}$ that in principle may correspond to any magnetic
texture and any type of SOC. 

After being transported over a closed Andreev trajectory that starts
at $x$ within the N region ($x_{L}<x<x_{R}$, $\Delta=0$), the BdG
bispinor should return to itself: 
\begin{equation}
\left(\begin{array}{c}
\psi\left(x\right)\\
\chi\left(x\right)
\end{array}\right)=e^{\mathbf{i}\left[2EL/v-2{\rm sgn}(v)\theta\left(E\right)+\varphi\right]}\left(\begin{array}{c}
W_{e}\psi\left(x\right)\\
W_{h}\chi\left(x\right)
\end{array}\right)\;,\label{eq:wf-loop}
\end{equation}
where $v=v_{F}\cos\phi$, and $\phi$ is the angle between the semiclassical
trajectory and the junction axis $x$. In Eq. (\ref{eq:wf-loop}),
the effect of spin-dependent fields is encoded in the electron and
hole spin rotation operators 
\begin{align}
W_{e}\left(x\right) & =U\left(x,x_{L}\right)\bar{U}\left(x_{L},x_{R}\right)U\left(x_{R},x\right)\quad,\label{eq:We-U}\\
W_{h}(x) & =\bar{U}\left(x,x_{R}\right)U\left(x_{R},x_{L}\right)\bar{U}\left(x_{L},x\right)\label{eq:Wh-U}
\end{align}
which are defined via the path-ordered spin propagator

\begin{equation}
U\left(x_{2},x_{1}\right)=\Pexp\left\{ -\dfrac{\mathbf{i}}{v}\int_{x_{1}}^{x_{2}}\boldsymbol{B}\left(v,x\right)\boldsymbol{\cdot\sigma}dx\right\} \;,\label{eq:U}
\end{equation}
 and its time-reversal conjugate $\bar{U}$ \footnote{Time-reversal conjugation of an operator $O\left({\bf p_{F}}\right)$
is given by $\bar{O}\left(\mathbf{p_{F}}\right)=\sigma^{y}O^{*}\left({\bf -p_{F}}\right)\sigma^{y}$ }.

The $W_{e,h}$ operators, which transport spinors over a closed trajectory,
are reminiscent of the Wilson loop operators in the SU(2) gauge theory.
They take into account the electron-hole conversions at the S/N interfaces
and they are thus defined along the Andreev loop. For this reason
we call $W_{e,h}$ the \textit{Andreev-Wilson }(AW)\textit{ loop operators},
which describe\textit{ }transport along a loop in the composite position
$\otimes$ particle-hole space \footnote{In the non-Abelian gauge theory a path-ordered exponential describing
a parallel transport of a spinor field over a closed loop in the real
space is called a Wilson operator, and its trace is called a Wilson
loop \cite{Peskin1995}. \eqref{eq:We-U} and \eqref{eq:Wh-U} also
represent the transport along a loop, now generated by the Andreev
reflections, and thus closes only in the combined position $\otimes$
particle-hole space. To recall this subtlety, we call $W_{e,h}$ the
\textit{Andreev-Wilson }(AW)\textit{ loop operators}.}. 

Several properties of $W_{e,h}$ are discussed in the supplemental
material \footnote{\label{fn:SM} See Supplemental Material at {[}URL will be inserted
by publisher{]} for a statement of the mathematical conventions followed
in this study, the mathematical structure behind Fig.\ref{fig:AW-loop},
a lengthy discussion of the detection of the Andreev-Wilson loop and
the derivation of Eq.\eqref{eq:ABS-T}.}. The most remarkable is that for any $\boldsymbol{B}(v,x)$ the trace
of $W_{e,h}\left(x\right)$ is $x$-independent, i.e. it does not
depend on the initial point of the loop. Hence $W_{e,h}\left(x\right)$
can be parametrized by local unit vectors $\boldsymbol{\mathfrak{n}}_{e,h}\left(x\right)$
and a coordinate independent angle $\Phi$:
\begin{equation}
W_{e,h}\left(x\right)=\exp\left[\mathbf{i}\left(\boldsymbol{\mathfrak{n}}_{e,h}\left(x\right)\boldsymbol{\cdot\sigma}\right)\Phi\right]\;.\label{eq:W-Phi-n}
\end{equation}
The vectors $\boldsymbol{\mathfrak{n}}_{e,h}(x)$ satisfy the classical
equation of a magnetic moment precessing in a magnetic field (see
SM): 

\begin{equation}
\pm v\partial_{x}\boldsymbol{\mathfrak{n}}_{e,h}=2\boldsymbol{B}\left(\pm v,x\right)\times\boldsymbol{\mathfrak{n}}_{e,h}\;.\label{eq:ne-nh}
\end{equation}
Since Andreev reflections preserve the spin, one has the boundary
condition $\boldsymbol{\mathfrak{n}}_{e}\left(x_{R,L}\right)=\boldsymbol{\mathfrak{n}}_{h}\left(x_{R,L}\right)$
{[}see (\ref{eq:We-U}-\ref{eq:Wh-U}){]} uniquely defining $\boldsymbol{\mathfrak{n}}_{e,h}(x)$.
One can easily see that expectation values of the electron and hole
spins, $\mathbf{s}_{e}(x)=\psi^{\dagger}\boldsymbol{\sigma}\psi$
and $\mathbf{s}_{h}(x)=\chi^{\dagger}\boldsymbol{\sigma}\chi$, have
$x$-independent projections on the local directions $\boldsymbol{\mathfrak{n}}_{e}(x)$
and $\boldsymbol{\mathfrak{n}}_{h}(x)$, respectively, see SM. Notice
that $\boldsymbol{\mathfrak{n}}_{e,h}$ and $\Phi$ depends on $v$.
From Eqs. (\ref{eq:We-U}-\ref{eq:Wh-U}) one can easily check that
$\boldsymbol{\mathfrak{n}}_{e}.\Phi|_{v}=-\boldsymbol{\mathfrak{n}}_{h}.\Phi|_{-v}$.
Based on the electron-hole symmetry we impose that $\boldsymbol{\mathfrak{n}}_{e}(v)=\boldsymbol{\mathfrak{n}}_{h}(-v)$.
From this follows that $\Phi(v)=-\Phi(-v)$. 

It is now possible to give a semiclassical interpretation of the AW-loops
(see Fig.\ref{fig:AW-loop}) inspired by the picture of quantization
for spinning particles proposed in Refs. \cite{Keppeler2002a,Keppeler2002}.
When transported along the loop the ``classical spins'' $\mathbf{s}_{e,h}(x)$
of electrons and holes precess around local axes, $\boldsymbol{\mathfrak{n}}_{e}(x)$
for the electrons and $\boldsymbol{\mathfrak{n}}_{h}(x)$ for the
holes, in such a way that latitude with respect to those axes is always
preserved. If, for example, one starts the AW-loop with a right moving
electron at position $x$, the electron spin will precess around the
local $\boldsymbol{\mathfrak{n}}_{e}$ until it reaches the right
electrode. At this point the electron is reflected as a hole. The
resulting hole propagates from the right to the left interface with
spin precessing around the local hole-like axis $\boldsymbol{\mathfrak{n}}_{h}$.
At the left interface the inverse process takes place, and the AW-loop
ends up with an electron precessing around $\boldsymbol{\mathfrak{n}}_{e}(x)$
axis again. While the rotation axis $\boldsymbol{\mathfrak{n}}_{e,h}$
after completing the loop is preserved, the spin itself does not return
to its original direction. There is an angle mismatch $\Phi$, at
fixed latitude with respect to $\boldsymbol{\mathfrak{n}}_{e}(x)$,
between the initial and final spin. Being position independent (since
$2\cos\Phi=\Tr\left\{ W_{e,h}\right\} $, see (\ref{eq:We-U}-\ref{eq:W-Phi-n}))
this angle mismatch has a global meaning: it corresponds to the phase
acquired by the wave-function after one turn. The single-valuedness
of the wave function after a complete period, expressed by Eq.\eqref{eq:wf-loop},
leads to the generalized semiclassical quantization condition,
\begin{equation}
\dfrac{2E_{n,s}L}{\left|v\right|}-2\arccos\dfrac{E_{n,s}}{\Delta}+{\rm sgn}\left(v\right)\left[\varphi+s\Phi\right]=2n\pi\label{eq:BS-spectrum}
\end{equation}
which determines the spectrum of ABS, with $s=\pm1$ being the spin
projection. The appearance of finite $\Phi$ lifts the spin degeneracy
of the ABS. From Eqs. (\ref{eq:We-U})-(\ref{eq:U}) one can verify
that the spin splitting occurs only if the effective magnetic field
$\boldsymbol{B}$ breaks the time-reversal symmetry, otherwise the
AW-loop operators are trivial. \footnote{The fact that only a non-zero Zeeman field leads to the spin splitting
of the ABS is a consequence of the leading order semiclassical approximation.
Beyond this approximation a lifting of the spin degenerancy can be
achieved only by SOC , see e.g. \cite{Dimitrova2006} if the phase
difference $\varphi$ is non-zero. This is directly linked to the
appearance of a finite magnetic moment by passing a current through
weak link with SOC \cite{Konschelle2015}.} 

As an example we consider the widely studied S/F/S junction (F is
a ferromagnet) \cite{buzdin.2005_RMP,Bergeret2005}. When the exchange
field points along the $z$-axis, $\boldsymbol{\mathfrak{n}}_{e}=\boldsymbol{\mathfrak{n}}_{h}=\hat{\boldsymbol{z}}$
are constant in space and $\Phi=2\int_{x_{L}}^{x_{R}}h_{z}\left(x\right)dx/v$.
In particular, this reduces to the usual $\Phi=2hL/v$ for a monodomain
S/F/S \cite{Bulaevskii1982,Buzdin1982}, and zero $\Phi$ for oscillating
exchange field with opposite domains of equal length \cite{Blanter2003}.
In fact, the previously known results for various specific S/F/S junctions
follow immediately from our general formulation that is valid for
arbitrary exchange field and SOC. The main message here is that all
the spin related features are encoded in the phase $\Phi$ and local
unit vectors $\boldsymbol{\mathfrak{n}}_{e,h}(x)$.

Clearly, the knowledge of the subgap spectral properties is not sufficient
to fully characterize the physics of the S/N/S junction. To understand,
for example, how the Josephson current is affected by the spin-dependent
interactions and finite temperature, or whether a finite magnetic
moment can be created in the junction one needs to extend the formalism
and take into account all energies in the spectrum and the electronic
distribution functions. For this we introduce the quasiclassical Green's
function, solve Eilenberger equation for the S/N/S junction and explore
how the vectors $\boldsymbol{\mathfrak{n}}_{e,h}$ and the phase $\Phi$,
associated with the AW-loop, manifest in physical observables. 

For a clean S/N/S junction in the presence of an effective magnetic
field $\boldsymbol{B}=\boldsymbol{h}+\boldsymbol{b}(v)$, where $\boldsymbol{h}$
is the spin-splitting/Zeeman field and $\boldsymbol{b}(-v)=-\boldsymbol{b}(v)$
describes SOC, the Eilenberger equation reads \cite{Bergeret2014,Konschelle2014}
\begin{equation}
\mathbf{i}v\partial_{x}\check{g}=\left[\tau_{3}E+\left(\tau_{3}\boldsymbol{h}+\boldsymbol{b}\right)\boldsymbol{\cdot\sigma}+\check{\Delta},\check{g}\left(x\right)\right]\;.\label{eq:transport-Heisenberg}
\end{equation}
Here $\check{g}$ is the matrix Green's function in the Nambu and
spin space, and $\check{\Delta}=\Delta e^{\mathbf{i}\tau_{3}\varphi/2}\mathbf{i}\tau_{2}e^{-\mathbf{i}\tau_{3}\varphi/2}$
is a 4$\times$4 matrix proportional to the identity matrix in spin
space and the Pauli matrices $\tau_{i}$ spanning the Nambu space.
We assume that $\Delta$ is constant and non-zero in S-electrodes
only, whereas $\boldsymbol{B}$ is present in the N-region. In Eq.
\eqref{eq:transport-Heisenberg} we only keep terms in the lowest
order in $(\xi p_{F})^{-1}$, where $\xi$ is any characteristic length
scale $\xi$ involved in the problem. Higher order terms are responsible
for the appearance of an anomalous phase in SFS structures and an
additional source for singlet-triplet conversion \cite{Konschelle2015,Reeg2015}.

By assuming continuity of the Green's functions across the interfaces
we obtain for the electron component $g$ of the Green's function
in N (the $(1,1)$ component of $\check{g}$ in Nambu space) \footnote{Details of the derivation will be given elsewhere, see also \cite{Konschelle2015}
for similar calculations.}:

\begin{equation}
g\left(x,E\right)=-\mathbf{i}\sum_{s=\pm}\frac{1}{2}\left(1+s\boldsymbol{\mathfrak{n}}_{e}\left(x\right)\boldsymbol{\cdot\sigma}\right)T_{s}\left(E\right)\;,\label{eq:g}
\end{equation}
with $s$ the spin-projection and 
\begin{equation}
T_{s}\left(E\right)=\tan\left(\dfrac{EL}{\left|v\right|}+\arcsin\dfrac{E}{\Delta}+{\rm sgn}(v)\left[\dfrac{\varphi}{2}+s\dfrac{\Phi}{2}\right]\right)\label{eq:T}
\end{equation}
The poles of $T_{s}$ for energies $E\leq\Delta$ represent the ABS,
and we thus recover Eq.\eqref{eq:BS-spectrum} explicitly. It is remarkable
that the precession angle mismatch $\Phi$ and the local spin precession
axis $\boldsymbol{\mathfrak{n}}_{e}(x)$ obtained from our previous
semiclassical consideration enter explicitly the Green's function.
We also note that factors $\frac{1}{2}\left(1\pm\boldsymbol{\mathfrak{n}}_{e}\left(x\right)\boldsymbol{\cdot\sigma}\right)=|\psi_{\pm}(x)\rangle\langle\psi_{\pm}(x)|$
in Eq.\eqref{eq:g} are exactly the projectors on the states with
spin up/down with respect to the local direction $\boldsymbol{\mathfrak{n}}_{e}\left(x\right)$.

The quasiclassical Green's function \eqref{eq:g} determines physical
observables like the density of states, given by 

\begin{equation}
\dfrac{N\left(E\right)}{N_{0}}=\frac{1}{\pi}\lim_{\epsilon\rightarrow0}\Re{\displaystyle \sum_{s=\pm}}\left\langle T_{s}\left(E+\mathbf{i}\epsilon\right)\right\rangle \label{eq:DOS}
\end{equation}
and the charge current through the junction $j=\left(1/4\right)\mathbf{i}\pi eN_{0}k_{B}T\sum_{\omega_{n}}\Tr\left\langle vg\right\rangle $,
where $\langle\dots\rangle$ denotes averaging over the Fermi surface
and the sum is over the Matsubara frequencies $\omega_{n}=2\pi k_{B}T\left(n+1/2\right)$.
After substitution of Eq.\eqref{eq:g} the charge current has the
form:

\begin{equation}
j=e\dfrac{\pi}{4}N_{0}k_{B}T\sum_{\omega_{n}}\sum_{s=\pm}\langle vT_{s}\left(\mathbf{i}\omega_{n}\right)\rangle\;.\label{eq:curr_density}
\end{equation}
This expression is valid for any Fermi surface, length of the junction,
magnetic interaction and the temperature. In that respect it generalizes
previous results obtained in ballistic S/F/S systems \cite{golubov_kupriyanov.2004,buzdin.2005_RMP,Bergeret2005,Konschelle2008}
to an arbitrary spin texture. Eq. \eqref{eq:curr_density} shows that
the current phase relation depends only on the parameter $\Phi$ irrespective
of its origin. For example, when $T$ is close to the critical temperature
$T_{c}$ the above expression simplifies to
\begin{equation}
\lim_{T\approx T_{c}}j=\dfrac{2eN_{0}\Delta^{2}}{\pi^{2}T_{c}}\left\langle \left|v\right|e^{-2\pi LT_{c}/\left|v\right|}\cos\Phi\right\rangle \sin\varphi\label{eq:current-Tc}
\end{equation}
which contains only the global magnetic phase shift $\Phi$, appearing
as a modulation of the Josephson current-phase relation. 

It is clear from Eqs. (\eqref{eq:DOS},\eqref{eq:curr_density}) that
spin-independent observables, such as the total density of states
or the charge current, do not depend on $\boldsymbol{\mathfrak{n}}_{e}$
and hence are constant in the N-region. In order to obtain information
about the vector $\boldsymbol{\mathfrak{n}}_{e}$, one needs to measure
spin-dependent observables. We introduce the spectral spin-density
polarized in $a$-direction (spin-resolved density of states) defined
by: 
\begin{equation}
\dfrac{N^{a}\left(E\right)}{N_{0}}=\lim_{\epsilon\rightarrow0}\dfrac{\Re\left\langle \Tr\left\{ \sigma^{a}g\right\} \right\rangle }{\pi}=\frac{1}{\pi}\Re{\displaystyle \sum_{s=\pm}}\left\langle s\boldsymbol{\mathfrak{n}}_{e}^{a}T_{s}\right\rangle \;,\label{eq:spin-DOS}
\end{equation}
that can be determined by means of tunneling spectroscopy similar
to the ABS spectroscopy done in carbon nanotubes connected to superconductors
\cite{Pillet2010}. If instead of nanotubes one uses semiconducting
wires with strong enough intrinsic SOC, a Zeeman field will induce
a finite $\Phi$ lifting the degeneracy of the ABS. The phase $\Phi$
will also manifest itself through the Josephson current according
to Eq.\eqref{eq:current-Tc}. In a similar experimental setup, one
can have access to the spectral spin density $N^{a}$ \eqref{eq:spin-DOS}.
Suppose the detector is fully-polarized (\textit{i.e.} in a half-metallic
limit) and magnetized along the $a$-direction, by performing two
measurements of the differential conductance for opposite magnetizations
of the tunneling probe, $G_{a}$ and $G_{-a}$ one determines $G_{a}\left(V\right)-G_{-a}\left(V\right)\propto\mathfrak{n}^{a}\sum_{s=\pm1}sT_{s}(V)$.
Thus, by measuring the total and spin-dependent density of states,
one can have an access to the parameters $\Phi$ and $\boldsymbol{\mathfrak{n}}_{e}$
which determine the full AW loop operator \eqref{eq:We-U}.

\begin{figure}[b]
\includegraphics[width=0.48\columnwidth]{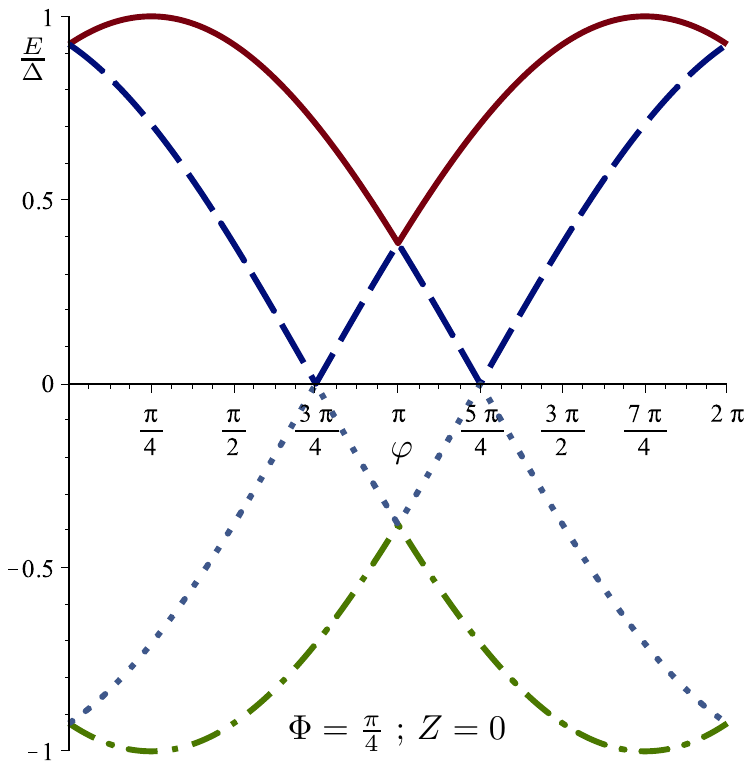}\includegraphics[width=0.48\columnwidth]{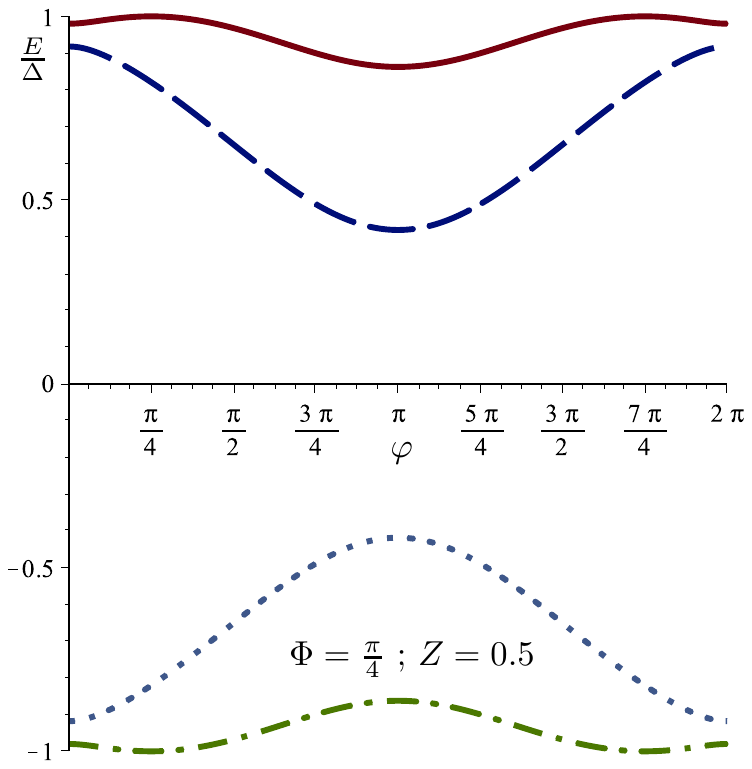}

\includegraphics[width=0.48\columnwidth]{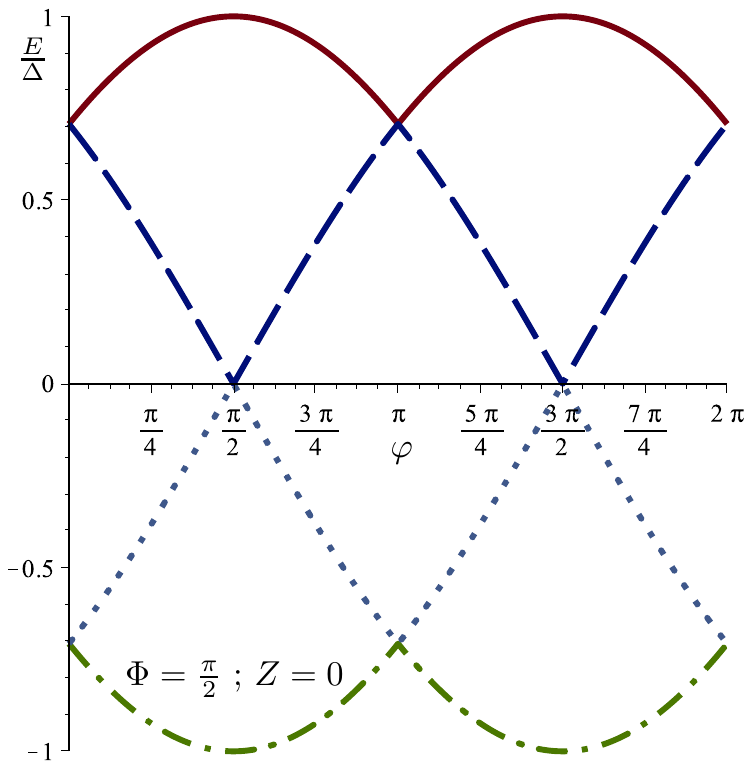}\includegraphics[width=0.48\columnwidth]{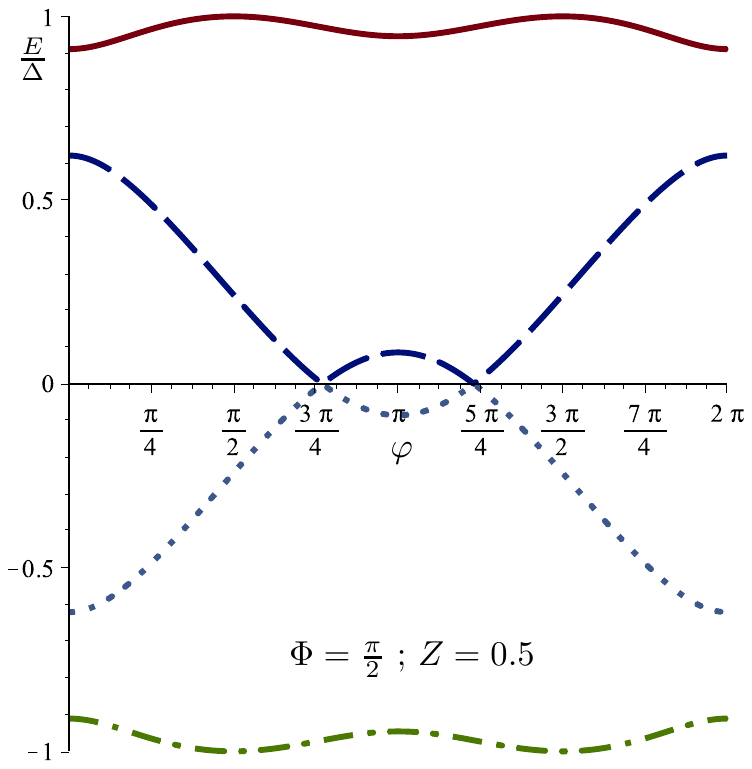}

\caption{\label{fig:ABS}The spin-splitted ABS energy $E/\Delta$ in a S/N/S
junction with a finite interface transparency versus the phase-difference
$\varphi$, in the limit $\Phi\gg2EL/v_{F}$ from Eq.\eqref{eq:ABS-T}.
Left column: the strength of the interface $\delta$-barrier $Z=0$,
right column: $Z=0.5$.}
\end{figure}

Previous results have been obtained assuming a perfect contact between
the S electrodes and the N link. One can however generalize the Bohr-Sommerfeld
quantization condition when adding scattering interfaces between the
S and N materials \cite{blonder_tinkham_klapwijk.1982}. Assuming
the left (L) and right (R) interfaces with transmission probabilities
$T_{L,R}=1-R_{L,R}$ and reflection coefficients $r_{L,R}$ with $\left|r_{L,R}\right|^{2}=R_{L,R}$
one obtains for a single channel junction (see SM):
\begin{multline}
\cos\left(\dfrac{2EL}{v_{F}}+s\Phi+2\theta\right)-\left(R_{R}+R_{L}\right)\cos\left(\dfrac{2EL}{v_{F}}+s\Phi\right)+\\
+R_{R}R_{L}\cos\left(\dfrac{2EL}{v_{F}}+s\Phi-2\theta\right)=\\
T_{L}T_{R}\cos\varphi-2\left(r_{L}r_{R}+r_{L}^{\ast}r_{R}^{\ast}\right)\sin^{2}\theta\label{eq:ABS-T}
\end{multline}
for the condition of existence of ABS. Eq.\eqref{eq:ABS-T} generalizes
results known for the case of S/N/S systems without magnetic interactions;
the case $\Phi=0$, $r_{L}=0$ has been obtained in \cite{bagwell.1992}.
Supposing a strong enough magnetic texture, such that $\Phi\gg2EL/v_{F}$,
one can plot the ABS of a short Josephson junction for different junction
transparency. As an example we consider a symmetric junction with
$\delta$-barriers at the interfaces: $r_{L}=r_{R}=-\mathbf{i}Z/\left(1+\mathbf{i}Z\right)$,
see Fig.\ref{fig:ABS}. In contrast to the case $\Phi=0$, where any
finite barrier strength ($Z\neq0)$ opens a gap, a finite $\Phi$
leads to a critical value of $Z$ below which zero-energy states exists.
As one can infer by comparing the upper and lower rows of Fig. \ref{fig:ABS},
this critical value of $Z$ increases by increasing $\Phi$.

In conclusion we have derived the semiclassical quantization condition
for a S/N/S Josephson junction when the normal region exhibits generic
SOC and exchange/Zeeman field. We obtained the spectrum of ABS Eq.\eqref{eq:BS-spectrum},
the quasiclassical Green's function Eq.\eqref{eq:g}, and analyzed
several physical observables in the presence of such generic spin-dependent
field. We demonstrated that all the properties of the junction are
expressed in terms of two fundamental parameters: $\Phi$ and $\boldsymbol{\mathfrak{n}}(x)$,
see Fig.\ref{fig:AW-loop}. These two parameters have a clear semiclassical
meaning. The unit vector $\boldsymbol{\mathfrak{n}}(x)$ describes
the local spin quantization axis about which a classical spin precess
at a constant latitude while propagating through the junction. The
magnetic phase $\Phi$ corresponds to the mismatch of the precession
angles after a quasiparticle completes the closed Andreev orbit. $\Phi$
enters explicitly the expression for the Josephson current, while
$\boldsymbol{\mathfrak{n}}$ can be accessed experimentally by measuring
the spin-resolved density of states. A magnetic Josephson junction
can thus be used as an analog computer of path-ordered loop operators
\eqref{eq:We-U}-\eqref{eq:W-Phi-n}. 
\begin{acknowledgments}
We thank D. Bercioux and V.N. Golovach for stimulating remarks. Discussions
with H. Bouchiat, P. Joyez, H. Pothier and L. Tosi during the annual
meeting of the GDR-I de Physique M\'{e}soscopique, Aussois 2015, were
particularly appreciated.

The work of F.S.B. and F.K. was supported by Spanish Ministerio de
Econom\'{\i}a y Competitividad (MINECO) through the Project No. FIS2014-55987-P
and the Basque Government under UPV/EHU Project No. IT-756-13. I.V.T.
acknowledges support from the Spanish Grant FIS2013-46159-C3-1-P,
and from the \textquotedblleft Grupos Consolidados UPV/EHU del Gobierno
Vasco\textquotedblright{} (Grant No. IT578-13)
\end{acknowledgments}

\appendix

\section{Convention}

The Green-Gor'kov functions read 
\begin{multline}
\check{G}\left(x_{1},x_{2};t_{1},t_{2}\right)=\dfrac{1}{\mathbf{i}\hbar}\left\langle \hat{T}\left[\left(\begin{array}{c}
\Psi\left(x_{1},t_{1}\right)\\
\left[\mathbf{i}\sigma_{2}\Psi\left(x_{1},t_{1}\right)\right]^{\dagger T}
\end{array}\right)\right.\right.\\
\left.\left.\otimes\left(\begin{array}{cc}
\Psi^{\dagger}\left(x_{2},t_{2}\right) & \left[\mathbf{i}\sigma_{2}\Psi\left(x_{2},t_{2}\right)\right]^{T}\end{array}\right)\right]\right\rangle 
\end{multline}
in the Nambu space, with 
\begin{equation}
\Psi\left(x,t\right)=\left(\begin{array}{c}
\Psi_{\uparrow}\\
\Psi_{\downarrow}
\end{array}\right)
\end{equation}
a spinor or annihilation operators of fermion with spin $\uparrow$
or $\downarrow$, respectively. $\Psi^{\dagger}$ is the creation
spinor associated to $\Psi$ and $^{T}$ represent the transpose.
$\hat{T}$ is the time-ordering operator, see \cite{Abrikosov1963}.

The quasi-classic Green's functions are defined as \cite{Langenberg1986}
\begin{equation}
\check{g}\left(x,t_{1},t_{2}\right)=\dfrac{\mathbf{i}}{\pi}\int d\xi\left[\check{G}\left(p,x,t_{1},t_{2}\right)\right]
\end{equation}
where $\xi=v_{F}\left(p-p_{F}\right)$ is the linearised increment
at the Fermi level, and $\check{G}\left(p,x\right)$ is obtained from
$\check{G}\left(x_{1},x_{2}\right)$ by the usual semi-classic expansion,
also known as Wigner transform or mixed coordinate Fourier transform
\cite{Langenberg1986}. Here we used a gauge-covariant generalisation
of it, the details of which can be found in \cite{Bergeret2014,Konschelle2014},
see also \cite{Gorini2010}. One then defines 
\begin{equation}
\check{g}=\left(\begin{array}{cc}
g & f\\
\bar{f} & -\overline{g}
\end{array}\right)
\end{equation}
in the Nambu space, with
\begin{equation}
g=\left(\begin{array}{cc}
g_{\uparrow\uparrow} & g_{\uparrow\downarrow}\\
g_{\downarrow\uparrow} & g_{\downarrow\downarrow}
\end{array}\right)\;;\;f=\left(\begin{array}{cc}
f_{\uparrow\downarrow} & f_{\uparrow\uparrow}\\
f_{\downarrow\downarrow} & f_{\downarrow\uparrow}
\end{array}\right)
\end{equation}
and so on for $\bar{f}=\mathcal{T}f\mathcal{T}^{-1}$ and $\bar{g}=\mathcal{T}g\mathcal{T}^{-1}$,
with $\mathcal{T}$ the time reversal operation. The components $g_{\alpha\beta}$,
$\alpha\in\left\{ \uparrow,\downarrow\right\} $ are correlation functions,
whereas $g$ and $f$ are matrices in the spin space (span by the
$\sigma$'s Pauli matrices), and $\check{g}$ is a matrix in the Nambu
space (span by the $\tau$'s Pauli matrices). In the main text we
calculate the matrix $g$ in the spin space, see Eq.(9).

\section{Andreev-Wilson loop\label{app:Andreev-Wilson-loop}}

Here we discuss basic properties of the Andreev-Wilson loop operators
defined by Eqs. (2)-(3) of the main text. The goal of this part is
to clarify the details of the semiclassical construction schematically
drawn on Fig.1 in the article.

Our starting point is the semiclassical Bogoliubov-deGennes (BdG)
equations (also known as Andreev approximation of the BdG equations)
\begin{align}
-\mathbf{i}\boldsymbol{v\cdot\nabla}\psi+\boldsymbol{B}\left(\boldsymbol{v},x\right)\boldsymbol{\cdot\sigma}\psi+\Delta\chi & =E\psi\nonumber \\
\mathbf{i}\boldsymbol{v\cdot\nabla}\chi+\boldsymbol{B}\left(-\boldsymbol{v},x\right)\boldsymbol{\cdot\sigma}\chi+\Delta^{\ast}\psi & =E\chi\label{eq:BdG}
\end{align}
for the spinor $\psi$ of the electron-like excitation and the spinor
$\chi$ for a hole-like excitation, with $\Delta$ being the superconducting
gap (assumed to be zero in the normal region), $\boldsymbol{B}$ a
magnetic texture which depends on the Fermi velocity $\boldsymbol{v}$
and position $x$, and $E$ the excitation energy. Here, we assume
that the system is translational invariant in the transversal direction
and therefore $\boldsymbol{v\cdot\nabla}\equiv v\partial_{x}$, where
the velocity $v=v_{F}\cos\phi,$ is parametrized by the angle between
the semiclassical trajectory and the $x$-axis perpendicular to the
S/N interfaces. 

Note that the equations \eqref{eq:BdG} correspond to the leading
order of the semi-classic expansion of the BdG equations in the presence
of any spin texture (spin splitting field plus spin-orbit coupling).

The operator $U\left(x_{2},x_{1}\right)$ defined in the main text
is the spin propagator that connects the values of the electron-like
spinor at two different points: $\psi\left(x_{2}\right)=e^{\mathbf{i}E\left(x_{2}-x_{1}\right)/v}U\left(x_{2},x_{1}\right)\psi\left(x_{1}\right)$.
From the BdG equations we find that $U\left(x_{2},x_{1}\right)$ satisfies
the following equations (in the N-region) 
\begin{align}
\mathbf{i}v\dfrac{\partial}{\partial x_{1}}U\left(x_{1},x_{2}\right) & =\boldsymbol{B}\left(v,x_{1}\right)\boldsymbol{\cdot\sigma}U\left(x_{1},x_{2}\right)\nonumber \\
\mathbf{i}v\dfrac{\partial}{\partial x_{2}}U\left(x_{1},x_{2}\right) & =-U\left(x_{1},x_{2}\right)\boldsymbol{B}\left(v,x_{1}\right)\boldsymbol{\cdot\sigma}\label{eq:Dyson-U}
\end{align}
with the boundary condition $U(x_{1},x_{1})=1$, and verifies the
group property $U\left(x_{1},x_{3}\right)U\left(x_{3},x_{2}\right)=U\left(x_{1},x_{2}\right)$.
The solution to these equations is given by the path-ordered exponential
\begin{equation}
U\left(x_{2},x_{1}\right)=\Pexp\left\{ -\dfrac{\mathbf{i}}{v}\int_{x_{1}}^{x_{2}}\boldsymbol{B}\left(v,x\right)\boldsymbol{\cdot\sigma}dx\right\} 
\end{equation}
which is the Eq.(4) of the main text.

The spin propagator for the hole-like spinor is defined similarly
as $\chi\left(x_{2}\right)=e^{-\mathbf{i}E\left(x_{2}-x_{1}\right)/v}\bar{U}\left(x_{2},x_{1}\right)\chi\left(x_{1}\right)$.
It follows from the BdG equations \eqref{eq:BdG} that operators $U$
and $\bar{U}$ are related via the time-reversal operation: $\bar{U}\left(v\right)=\sigma^{y}U^{\ast}\left(-v\right)\sigma^{y}$.
Explicitly the path-ordered exponential representation for $\bar{U}\left(x_{2},x_{1}\right)$
reads 
\begin{equation}
\bar{U}\left(x_{2},x_{1}\right)=\Pexp\left\{ \dfrac{\mathbf{i}}{v}\int_{x_{1}}^{x_{2}}\boldsymbol{B}\left(-v,x\right)\boldsymbol{\cdot\sigma}dx\right\} .
\end{equation}
Note that $U$ can be also obtained from $\bar{U}$ by simply reversing
the velocity $\bar{U}(v)=U(-v)$, which reflects the electron-hole
symmetry.

From the spin propagators $U$ and $\bar{U}$ we construct the Andreev-Wilson
loop operators {[}Eqs.(2) and (3) of the main text{]}
\begin{align}
W_{e}\left(x\right) & =U\left(x,x_{L}\right)\bar{U}\left(x_{L},x_{R}\right)U\left(x_{R},x\right)\label{eq:We-U-1}\\
W_{h}(x) & =\bar{U}\left(x,x_{R}\right)U\left(x_{R},x_{L}\right)\bar{U}\left(x_{L},x\right)\:.\label{eq:Wh-U-1}
\end{align}
These operators propagate electron-like and hole-like spinors in the
particle-hole $\otimes$ position space along an Andreev loop between
the two superconducting electrodes at locations $x_{L}$ and $x_{R}$
{[}see Fig.1, main text{]}. For example the operator $W_{e}(x)$ propagates
the electron spinor from a point $x$ to the right interface at $x_{R}$,
then the hole spinor from $x_{R}$ to the left interface at $x_{L}$,
and finally it transfers the electron spinor from $x_{L}$ back to
the original point $ $$x$. 

Since the above defined Andreev-Wilson loop operators are $\text{SU}\left(2\right)$
rotation matrices we can represent them in the following form 
\begin{align}
W_{e,h}\left(x\right) & =\exp\left[\mathbf{i}\left(\boldsymbol{\mathfrak{n}}_{e,h}\left(x\right)\boldsymbol{\cdot\sigma}\right)\Phi\right]\label{eq:W-Phi-n-1}\\
 & =\cos\Phi+\mathbf{i}\left(\boldsymbol{\mathfrak{n}}_{e,h}\left(x\right)\boldsymbol{\cdot\sigma}\right)\sin\Phi\,,\label{eq:W-cos-sin}
\end{align}
where $\boldsymbol{\mathfrak{n}}_{e}\left(x\right)$ and $\boldsymbol{\mathfrak{n}}_{h}\left(x\right)$
are unit vectors. Using the group property of the propagators $U(x_{1},x_{2})$
and $\bar{U}(x_{1},x_{2})$ we find the relation 
\begin{equation}
\Tr\left\{ W_{e,h}\right\} =\Tr\left\{ U\left(x_{R},x_{L}\right)\bar{U}\left(x_{L},x_{R}\right)\right\} =2\cos\Phi\label{eq:Phi-position-independent}
\end{equation}
which shows that the parameter $\Phi$ is $x$-independent and the
same for the electron- and hole-like loop operators. In contrast to
that the vectors $\boldsymbol{\mathfrak{n}}_{e}\left(x\right)$ and
$\boldsymbol{\mathfrak{n}}_{h}\left(x\right)$ are in general different
and $x$-dependent. The physical significance of the parametrization
\eqref{eq:W-Phi-n-1} is the following. When the expectation values
of the electron $\mathbf{s}_{e}\left(x\right)=\psi^{\dagger}\left(x\right)\boldsymbol{\sigma}\psi\left(x\right)$
and hole $\mathbf{s}_{h}\left(x\right)=\chi^{\dagger}\left(x\right)\boldsymbol{\sigma}\chi\left(x\right)$
spin vectors are propagated around the Andreev-Wilson loops based
at the point $x$, they rotate by the angle $\Phi$ about the directions
of $\boldsymbol{\mathfrak{n}}_{e}\left(x\right)$ and $\boldsymbol{\mathfrak{n}}_{h}\left(x\right)$,
respectively. 

Using \eqref{eq:Dyson-U} and the definitions \eqref{eq:We-U-1}-\eqref{eq:Wh-U-1}
we find the equations of motion for the Andreev-Wilson loop operators,
\begin{multline}
\mathbf{i}v\dfrac{\partial}{\partial x}W_{e}\left(x\right)=\mathbf{i}v\dfrac{\partial U\left(x,x_{L}\right)}{\partial x}\bar{U}\left(x_{L},x_{R}\right)U\left(x_{R},x\right)\\
+\mathbf{i}vU\left(x,x_{L}\right)\bar{U}\left(x_{L},x_{R}\right)\dfrac{\partial U\left(x_{R},x\right)}{\partial x}\\
=\left[\boldsymbol{B}\left(v,x\right)\boldsymbol{\cdot\sigma},W_{e}\left(x\right)\right]\label{eq:Heisenberg-W}
\end{multline}
and similarly for $W_{h}(x)$: 
\begin{equation}
-\mathbf{i}v\partial_{x}W_{h}\left(x\right)=\left[\boldsymbol{B}\left(-v,x\right)\boldsymbol{\cdot\sigma},W_{h}\left(x\right)\right]\,.
\end{equation}
By substituting the representation \eqref{eq:W-cos-sin} into the
relation \eqref{eq:Heisenberg-W} and recalling the property \eqref{eq:Phi-position-independent},
one gets 
\begin{equation}
\pm\mathbf{i}v\partial_{x}\left(\boldsymbol{\mathfrak{n}}_{e,h}\left(x\right)\boldsymbol{\cdot\sigma}\right)=\left[\boldsymbol{B}\left(\pm v,x\right)\boldsymbol{\cdot\sigma},\boldsymbol{\mathfrak{n}}_{e,h}\left(x\right)\boldsymbol{\cdot\sigma}\right]
\end{equation}
Next, using the identity $\left(\boldsymbol{A\cdot\sigma}\right)\left(\boldsymbol{B\cdot\sigma}\right)=\boldsymbol{A\cdot B}+\mathbf{i}\left(\boldsymbol{A\times B}\right)\boldsymbol{\cdot\sigma}$
for $\boldsymbol{A}$ and $\boldsymbol{B}$ in $\mathbb{R}^{3}$ to
evaluate the commutator, we eventually arrive at the following classical
equations of precession of the vectors $\boldsymbol{\mathfrak{n}}_{e,h}$
around the effective magnetic field $\boldsymbol{B}$: 
\begin{align}
v\partial_{x}\boldsymbol{\mathfrak{n}}_{e} & =2\boldsymbol{B}\left(v,x\right)\times\boldsymbol{\mathfrak{n}}_{e}\nonumber \\
v\partial_{x}\boldsymbol{\mathfrak{n}}_{h} & =-2\boldsymbol{B}\left(-v,x\right)\times\boldsymbol{\mathfrak{n}}_{h}\label{eq:ne-nh-1}
\end{align}
This corresponds to Eq.(6) of the main text. 

To establish the boundary conditions for Eqs.\eqref{eq:ne-nh-1} we
evaluate the loop operators \eqref{eq:We-U-1}-\eqref{eq:Wh-U-1}
at the interface points 
\begin{equation}
W_{e}\left(x_{L}\right)=\bar{U}\left(x_{L},x_{R}\right)U\left(x_{R},x_{L}\right)=W_{h}\left(x_{L}\right)
\end{equation}
\begin{equation}
W_{h}\left(x_{R}\right)=U\left(x_{R},x_{L}\right)\bar{U}\left(x_{L},x_{R}\right)=W_{e}\left(x_{R}\right)
\end{equation}
These relations imply that $\boldsymbol{\mathfrak{n}}_{e}\left(x_{L,R}\right)=\boldsymbol{\mathfrak{n}}_{h}\left(x_{L,R}\right)$
at the interfaces with the superconducting electrodes, since $\Phi$
is position independent, see \eqref{eq:Phi-position-independent}.
Therefore, despite the Bogoliubov quasiparticles bouncing back at
the S/N interfaces, $x=x_{L,R}$ and transmuting there (electron-like
excitation becomes a hole-like excitation and vice-versa by the Andreev
reflection), the local precession vectors $\boldsymbol{\mathfrak{n}}_{e}$
and $\boldsymbol{\mathfrak{n}}_{h}$ are equal at the interfaces,
hence the propagation of the spinors $\psi$ and $\chi$ can be defined
continuously along an Andreev loop.

Obviously the electron $\mathbf{s}_{e}\left(x\right)=\psi^{\dagger}\left(x\right)\boldsymbol{\sigma}\psi\left(x\right)$
and hole spin vector $\mathbf{s}_{h}\left(x\right)=\chi^{\dagger}\left(x\right)\boldsymbol{\sigma}\chi\left(x\right)$,
also satisfy the precession equations similar to Eqs.\eqref{eq:ne-nh-1}.
Hence one naturally expects that the scalar product $\mathbf{s}_{e,h}(x)\boldsymbol{\cdot\mathfrak{n}}_{e,h}(x)$
should be preserved along the loop. Indeed, using the BdG equations
\eqref{eq:BdG} and the relation \eqref{eq:ne-nh-1} one has 
\begin{multline}
\mathbf{i}v\dfrac{\partial\mathbf{s}_{e}\boldsymbol{\cdot\mathfrak{n}}_{e}}{\partial x}=\mathbf{i}v\dfrac{\partial\psi^{\dagger}}{\partial x}\boldsymbol{\sigma\cdot\mathfrak{n}}_{e}\psi+\mathbf{i}v\psi^{\dagger}\dfrac{\partial\boldsymbol{\sigma\cdot\mathfrak{n}}_{e}}{\partial x}\psi\\
+\mathbf{i}v\psi^{\dagger}\boldsymbol{\sigma\cdot\mathfrak{n}}_{e}\dfrac{\partial\psi}{\partial x}=-\psi^{\dagger}\left(\boldsymbol{B}\left(v\right)\boldsymbol{\cdot\sigma}\right)\left(\boldsymbol{\mathfrak{n}}_{e}\boldsymbol{\cdot\sigma}\right)\psi\\
+\psi^{\dagger}2\mathbf{i}\left(\boldsymbol{B}\left(v\right)\boldsymbol{\times\mathfrak{n}}_{e}\right)\boldsymbol{\cdot\sigma}\psi+\psi^{\dagger}\left(\boldsymbol{\mathfrak{n}}_{e}\boldsymbol{\cdot\sigma}\right)\left(\boldsymbol{B}\left(v\right)\boldsymbol{\cdot\sigma}\right)\psi\\
\Rightarrow\mathbf{i}v\dfrac{\partial\mathbf{s}_{e}\boldsymbol{\cdot\mathfrak{n}}_{e}}{\partial x}=0\;.
\end{multline}
Here we have used that $\left(\boldsymbol{A\cdot\sigma}\right)\left(\boldsymbol{B\cdot\sigma}\right)=\boldsymbol{A\cdot B}+\mathbf{i}\left(\boldsymbol{A\times B}\right)\boldsymbol{\cdot\sigma}$
for $\boldsymbol{A}$ and $\boldsymbol{B}$ in $\mathbb{R}^{3}$.
Similarly we find that $\mathbf{s}_{h}\boldsymbol{\cdot\mathfrak{n}}_{h}$
is also space independent. Thus the projection of the expectation
value $\mathbf{s}_{e,h}(x)$ of the electron/hole spin on the local
axis $\boldsymbol{\mathfrak{n}}_{e,h}(x)$ remains constant in the
course of propagation along the closed Andreev trajectory. In other
words, the spin dynamics can be viewed as a precession at a constant
latitude about a local axis that itself changes along the trajectories
according to Eq.\eqref{eq:ne-nh-1}. All this confirms the picture
illustrated in Fig.1 in the main text.

Finally we note that, mathematically speaking, the constancy of the
projection of the spin vector $\mathbf{s}_{e,h}(x)$ on the local
axis $\boldsymbol{\mathfrak{n}}_{e,h}(x)$ means the existence of
an extra integral of motion in the classical spin dynamics. This additional
integral of motion allows to reduce the SU(2) holonomy (expected in
the spin-1/2 problem) to an U(1) holomony parametrized by a single
scalar parameter $\Phi$. A similar property has been identified by
Keppeler in the context of semilassical quantization of spinning electrons
described by Pauli or Dirac equations \cite{Keppeler2002,Keppeler2002a}.
Our results in fact show that the closed Andreev trajectory can be
interpreted as a special case of the generalized invariant torus introduced
in Refs.\cite{Keppeler2002,Keppeler2002a}. The important difference
is however that in our case $\Phi$ is the holonomy associated to
the equations \eqref{eq:Dyson-U} along the path shown in Fig.1 (main
text) mixing electron and hole trajectories. The non-integrable loop
so formed exists in the particle-hole $\otimes$ position space.

\section{Josephson junction as an analog computer of path-ordered exponential}

As discussed above, the Andreev-Wilson loop operators describe the
spin dependency of the BdG bispinor \eqref{eq:BdG} as they move along
the normal region. The loop can not be defined only in the position
space since this one is purely 1D along a quasi-classic trajectory,
but in the particle-hole $\otimes$ position space. It is clear that
without a time-reversal breaking field in the normal region, one has
trivial operators $W_{e}=W_{h}=1$ and no $\Phi$-holonomy. There
is in fact no loop in this case. So the surface of the loop is described
by the amplitude of the time-reversal breaking field (which gives
the difference between $U$ and $\bar{U}$) and the length of the
junction. To recall this subtlety, we coined $W_{e,h}$ the \textit{Andreev}-Wilson
loop operators. They are \textit{not} Wilson loop in the usual sense.

As described in the previous section of this supplemental materials,
the Andreev-Wilson loop operators follow the usual equation of motion
for the quantum state, providing we replace the usual time evolution
-- the Schr\"{o}dinger equation -- by the transport equation \eqref{eq:Dyson-U}
for the evolution / displacement operator $U$. 

An important problem for quantum information is to understand the
time evolution of the qubit state. This evolution is defined as a
time-ordered Dyson series, which usually require heavy computational
power to be calculated, even in a perturbative way. We have shown
in the main text that a Josephson junction with magnetic interaction
is an analog computer for such complicated mathematical objects. In
fact, the Andreev-Wilson operator can be completely described by the
two parameters $\Phi$ and $\boldsymbol{\mathfrak{n}}_{e,h}$, and
one can access these two parameters by measuring the current-phase
relation and/or the density of state and the spin polarized density
of states, see the main text. Extracting $\Phi$ and $\boldsymbol{\mathfrak{n}}_{e,h}$
from such measurements, one can measure the result of the Dyson equations
\eqref{eq:Dyson-U} in an analog fashion. 

The Josephson devices might be versatile enough, since either the
spin-orbit or the spin-splitting interaction could be tuned via external
gates voltages, mutual inductances, etc. In addition, the quantity
$L/v$ plays in the spinor propagation of the S/N/S junction the role
of the time in the usual Schr\"{o}dinger equation ; $L/v$ can be
tuned as well (though with more difficulties than the magnetic texture
itself). 

Note finally that the expression for the quasiclassical Green's function
(Eqs.(9-10) of the main text) is pretty generic: if instead of a spin
field, one considers any other Lie group structure associated to the
generator of the Lie algebra $\mathfrak{s}$, the generator will simply
appear by the replacement $\boldsymbol{\sigma}\rightarrow\boldsymbol{\mathfrak{s}}$
in the expression for the Green function, as long as this algebra
will commute with the Nambu structure. In particular, a generalization
to the recently proposed SU(N) cold atomic gas is straightforward
\cite{Banerjee2013}.

Thus, in certain sense our study is just a first step toward a possible
understanding of the complete analogy between the transport of electron
spin in coherent structures and the time evolution of complex quantum
systems.

\section{Junction with barriers}

In this section, we demonstrate the general expression (16) of the
main text for the spectrum of the Andreev bound states when the junction
has some barriers. We first consider the scattering of particles and
holes with $E<\Delta$ at a single interface between a normal metal
(left half space) and a superconductor (right half space). Then we
discuss the case of two interfaces in order to get the spectrum of
the Andreev bound states. All this analysis is done for a single-channel
junction.

Notice there are other approaches to avoid discussing pure ballistic
transport in Josephson junction. Perhaps the simplest one we could
think of would be to adopt a diffusive equation approach \cite{Bergeret2014}.
This is let for future study. Other ways are to add a scattering region
in the normal part of the junction \cite{Beenakker1991}, or to add
a barrier in the middle of the junction \cite{bagwell.1992}. Nevertheless,
these approaches are not easy to handle in our case, since one would
like to preserve the Andreev-Wilson loop. So we adopt the method of
adding impurities only at the S/N and N/S interfaces.

\subsection{Single interface problem}

Let us thus model a finite transparency interface by a potential barrier
in the normal region at the distance from the interface $l\ll\xi_{0}$
smaller than the coherence length. The reason is due to the impossibility
to treat a microscopic barrier in the semi-classic approximation.
One has to add a normal region at the distance $l\ll\xi_{0}$ where
microscopic theory should be prefered, and recover the semi-classic
limit at large scale $\xi_{0}$. The BdG bispinor reads
\begin{equation}
\left(\begin{array}{c}
u\left(x\right)\\
v\left(x\right)
\end{array}\right)=\left(\begin{array}{c}
\psi_{+}\left(x\right)\\
\chi_{+}\left(x\right)
\end{array}\right)e^{\mathbf{i}\mathbf{k\cdot r}}+\left(\begin{array}{c}
\psi_{-}\left(x\right)\\
\chi_{-}\left(x\right)
\end{array}\right)e^{-\mathbf{i}\mathbf{k\cdot r}}\label{BdGfunctions}
\end{equation}
where the slowly varying functions $\psi_{\pm}$ and $\chi_{\pm}$
are the electron and hole spinors for two Fermi points labeled by
the indexes $+$ and $-$. Note that the states $\psi_{+}$ and $\chi_{-}$
correspond to the right moving quasiparticles (positive velocity),
while $\psi_{-}$ and $\chi_{+}$ describe the left movers (negative
velocity). The barrier in the N-region is parametrized by the transmission
$t$ and the reflection $r$ coefficients, which satisfy the following
identities
\begin{eqnarray}
\left|t\right|^{2}+\left|r\right|^{2} & = & 1\label{identity-1}\\
tr^{*}+t^{*}r & = & 0\label{identity-2}
\end{eqnarray}
It is convenient to introduce the electron and hole bispinors
\begin{equation}
\Psi=\left(\begin{array}{c}
\psi_{+}\\
\psi_{-}
\end{array}\right)\;\text{and}\;\Phi=\left(\begin{array}{c}
\chi_{+}\\
\chi_{-}
\end{array}\right)\label{Psi-Phi}
\end{equation}
and the transmission and reflection matrices
\begin{equation}
\hat{\mathbf{t}}=\left(\begin{array}{cc}
t & 0\\
0 & t^{*}
\end{array}\right)\;\text{and}\;\hat{\textbf{r}}=\left(\begin{array}{cc}
0 & r\\
r^{*} & 0
\end{array}\right)\label{t-r-matrices}
\end{equation}
which satisfy the relations
\begin{equation}
\hat{\mathbf{t}}\hat{\mathbf{t}}^{\ast}=\left|t\right|^{2},\quad\hat{\textbf{r}}^{2}=\left|r\right|^{2},\quad\hat{\textbf{r}}\hat{\mathbf{t}}=-\hat{\mathbf{t}}\hat{\textbf{r}}^{\ast}\label{eq:matrix-ident}
\end{equation}
Importantly, the barrier scattering is the same for the electrons
and holes as this is a Fermi surface property related to the fast
oscillating parts of the BdG bispinors.

Then the scattering relation between the states on the right (labeled
$^{R}$) and on the left (labeled $^{L}$) sides of the barrier can
be written as follows. 
\begin{eqnarray}
\Psi^{R} & = & \hat{\textbf{r}}\Psi^{R}+\hat{\mathbf{t}}\Psi^{L}\label{e-scatt}\\
\Phi^{R} & = & \hat{\textbf{r}}\Phi^{R}+\hat{\mathbf{t}}\Phi^{L}\label{h-scatt}
\end{eqnarray}
From these relations we can construct the left-to-right transfer matrix
defined as $\Psi^{R}=T\Psi^{L}$
\begin{align}
\Psi^{R} & =\left(1-\hat{\textbf{r}}\right)^{-1}\hat{\mathbf{t}}\Psi^{L}\nonumber \\
 & =\left(1+\hat{\textbf{r}}\right)\frac{\hat{\mathbf{t}}}{\left|t\right|^{2}}\Psi^{L}=\frac{\hat{\mathbf{t}}}{\left|t\right|^{2}}\left(1-\hat{\textbf{r}}^{\ast}\right)\Psi^{L}
\end{align}
using the matrix identities \eqref{eq:matrix-ident}. Therefore the
transfer matrix is identified as
\begin{equation}
T=(1+\hat{\textbf{r}})\frac{\hat{\mathbf{t}}}{\left|t\right|^{2}}=\frac{\hat{\mathbf{t}}}{\left|t\right|^{2}}(1-\hat{\textbf{r}}^{\ast})\label{eq:transfer-matrix}
\end{equation}
Using Eqs.(\ref{eq:matrix-ident}) and (\ref{eq:transfer-matrix})
one can check that $TT^{*}=1$. Therefore the right-to-left transfer
is described by $T^{*},$ that is $\Psi^{L}=T^{*}\Psi^{R}$.

The transfer matrices for the hole states are given by the same formulas.

The scattering of the states located between the barrier and the S-region
(the states $\Psi^{R}$ and $\Phi^{R}$ in our case) is the pure Andreev
scattering at the ideal fully transparent N/S interface
\begin{equation}
\Phi^{R}=S_{0}\Psi^{R},\;S_{0}=e^{-\mathbf{i}\varphi}\left(\begin{array}{cc}
e^{-\mathbf{i}\theta} & 0\\
0 & e^{\mathbf{i}\theta}
\end{array}\right)\label{Andreev-0}
\end{equation}
where $\varphi$ is the SC phase and $\theta=\arccos\frac{E}{\Delta}$,
see e.g. \cite{Beenakker1991}. 

In the presence of the barrier the generalized Andreev scattering
from the electrons to the holes on the left-hand-side of the barrier,
$\Psi^{L}\to\Phi^{L}$, is composed from the two transfers across
the barrier (forward L-to-R for electrons, and back R-to-L for holes)
and a pure Andreev scattering in between: $\Psi^{L}\to\Psi^{R}\to\Phi^{R}\to\Phi^{L}$.
This process is described as follows
\begin{equation}
\Phi^{L}=T^{*}S_{0}T\Psi^{L}\equiv A\Psi^{L}\label{Andreev-1}
\end{equation}
Explicitly for the generalized Andreev scattering matrix we get
\begin{eqnarray}
A & = & (1+\hat{\mathbf{r}}^{\ast})\frac{\hat{\mathbf{t}}^{\ast}}{\left|t\right|^{2}}S_{0}\frac{\hat{\mathbf{t}}}{\left|t\right|^{2}}(1-\hat{\mathbf{r}}^{\ast})=(1+\hat{\mathbf{r}}^{\ast})\frac{S_{0}}{\left|t\right|^{2}}(1-\hat{\mathbf{r}}^{\ast})\nonumber \\
 & = & \frac{e^{-\mathbf{i}\varphi}}{\left|t\right|^{2}}\left(\begin{array}{cc}
e^{-\mathbf{i}\theta}-\left|r\right|^{2}e^{\mathbf{i}\theta} & r^{*}\left(e^{\mathbf{i}\theta}-e^{-\mathbf{i}\theta}\right)\\
r\left(e^{-\mathbf{i}\theta}-e^{\mathbf{i}\theta}\right) & e^{\mathbf{i}\theta}-\left|r\right|^{2}e^{-\mathbf{i}\theta}
\end{array}\right)\label{Andreev-2}
\end{eqnarray}
The matrix $A$ connects the electron and hole states scattered by
a non ideal N/S interface modeled by the barrier and an ideal N/S
interface. In the following we will skip the index $^{L}$, since
the solutions are now all in the left/normal region, thus we have:
\begin{multline}
\left(\begin{array}{c}
\chi_{+}\\
\chi_{-}
\end{array}\right)=\\
\frac{e^{-\mathbf{i}\varphi}}{\left|t\right|^{2}}\left(\begin{array}{cc}
e^{-\mathbf{i}\theta}-\left|r\right|^{2}e^{\mathbf{i}\theta} & r^{*}\left(e^{\mathbf{i}\theta}-e^{-\mathbf{i}\theta}\right)\\
r\left(e^{-\mathbf{i}\theta}-e^{\mathbf{i}\theta}\right) & e^{\mathbf{i}\theta}-\left|r\right|^{2}e^{-\mathbf{i}\theta}
\end{array}\right)\left(\begin{array}{c}
\psi_{+}\\
\psi_{-}
\end{array}\right)\label{Andreev-3}
\end{multline}
As the final step one can construct the interface scattering matrix
$S$ which connects the right moving incident (electron and hole in
the N region) states to the left moving reflected (electron and hole
in the N region) states:
\begin{equation}
\left(\begin{array}{c}
\psi_{-}\\
\chi_{+}
\end{array}\right)=S\left(\begin{array}{c}
\psi_{+}\\
\chi_{-}
\end{array}\right)\label{S-def}
\end{equation}
The elements of the interface $S$ matrix are constructed from Eqs.(\ref{Andreev-3})
by expressing $\psi_{-}$ and $\chi_{+}$ in terms of $\psi_{+}$
and $\chi_{-}$. The resulting scattering matrix takes the form
\begin{multline}
S=\frac{1}{e^{\mathbf{i}\theta}-\left|r\right|^{2}e^{-\mathbf{i}\theta}}\left(\begin{array}{cc}
r\left(e^{\mathbf{i}\theta}-e^{-\mathbf{i}\theta}\right) & \left|t\right|^{2}e^{\mathbf{i}\varphi}\\
\left|t\right|^{2}e^{-\mathbf{i}\varphi} & r^{*}\left(e^{\mathbf{i}\theta}-e^{-\mathbf{i}\theta}\right)
\end{array}\right)\\
\equiv\left(\begin{array}{cc}
r_{e} & r_{A}e^{\mathbf{i}\varphi}\\
r_{A}e^{-\mathbf{i}\varphi} & r_{h}
\end{array}\right)\label{S-result}
\end{multline}
where we introduced the notations $r_{e}$ , $r_{h}$ for the electron
and hole normal reflection coefficients and $r_{A}$ for the Andreev
(electron-hole) reflection coefficient
\begin{eqnarray}
r_{A} & = & \frac{\left|t\right|^{2}}{e^{\mathbf{i}\theta}-\left|r\right|^{2}e^{-\mathbf{i}\theta}}\label{rA}\\
r_{e} & = & r\frac{e^{\mathbf{i}\theta}-e^{-\mathbf{i}\theta}}{e^{\mathbf{i}\theta}-\left|r\right|^{2}e^{-\mathbf{i}\theta}}\label{re}\\
r_{h} & = & r^{*}\frac{e^{\mathbf{i}\theta}-e^{-\mathbf{i}\theta}}{e^{\mathbf{i}\theta}-\left|r\right|^{2}e^{-\mathbf{i}\theta}}\label{rh}
\end{eqnarray}
One can verify that $\left|r_{A}\right|^{2}+\left|r_{e,h}\right|^{2}=1$,
and the interface $S$-matrix is unitary $S^{\dagger}S=1.$ 

It seems to be quite obvious that in the case of an S/N interface
(S on the left, N on the right), the same S-matrix (\ref{S-result})
connects the left moving (incoming) states $\psi_{-}$ and $\chi_{+}$
to the right moving (reflected) states $\psi_{+}$ and $\chi_{-}$
(the incoming and outgoing are interchanged for S/N with respect to
N/S).

\subsection{Andreev bound states spectrum}

In the Andreev bound states problem we have two interfaces (L and
R) with $S$-matrices $S_{R}$ and $S_{L}$ given by Eq.(\ref{S-result})
where the barrier transmission/reflection and the SC phases are in
general different for L and R interfaces. The propagation between
interfaces is diagonal in the electron-hole space (the same for either
left movers and time-conjugated for right movers) and can be diagonalized
locally in the spin space. As a result the single-valuedness condition
(for the loop starting at the left boundary) reduces to the following
simple form:
\begin{equation}
S_{L}S_{R}e^{\mathbf{i}\left(\frac{2E}{v}L+s\Phi\right)}\left(\begin{array}{c}
\psi_{+}(x_{L})\\
\chi_{-}(x_{L})
\end{array}\right)=\left(\begin{array}{c}
\psi_{+}(x_{L})\\
\chi_{-}(x_{L})
\end{array}\right)\label{singlevaludness}
\end{equation}
Thus the problem reduces to a simple $2\times2$ system (for each
spin projection $s=\pm1$) with a very clear physical meaning -- the
phase accumulated in the free closed loop propagation should be compensated
by the combined scattering on two interfaces. The matrix in the r.h.s.
is the generalization of our Andreev-Wilson loop to the case of general
interfaces with finite transparency.

To obtain the spectrum of the Andreev bound states, one should thus
resolve the condition
\begin{equation}
\det\left(1-S_{L}S_{R}e^{\mathbf{i}\left(\frac{2E}{v}L+s\Phi\right)}\right)=0
\end{equation}
in order to get non-trivial solution verifying \eqref{singlevaludness}.
After a straightforward calculation of the determinant and some algebra
one obtains Eq.(16) of the main text. 

Eq.(16) is quite generic since it is valid for any interface with
all properties encoded in the expressions for $r$ and $t$ and any
spin texture, encoded in the $\Phi$-holonomy. Eq.(16) thus generalizes
many results scattered in the existing literature about single-channel
Josephson junction, either for the S/N/S or the S/F/S problems.

\end{document}